 \documentclass[aps,prl,twocolumn,groupedaddress,amsmath,amssymb,showpacs]{revtex4}
\usepackage{graphicx}
\usepackage{amsmath}

\begin{document}

\title{Ray chaos in optical cavities based upon standard laser mirrors}

%\title{Optical Cavities as ``Open'' Billiards: Evidence for Ray Chaos }

\author{A. Aiello}
\author{M. P. van Exter}
\author{J. P. Woerdman}
\affiliation{Huygens Laboratory, Leiden University, P.O. Box 9504,
Leiden, The Netherlands}
\date{\today}
\begin{abstract}
We present a composite optical cavity made  of standard laser
mirrors; the cavity consists  of a suitable combination of stable
and unstable cavities. In spite of its very open nature the
composite cavity shows ray chaos, which may be either soft or
hard, depending on the cavity configuration. This opens a new,
convenient route for experimental studies of the quantum aspects
of a chaotic wave field.
\end{abstract}
\maketitle
%
%\baselineskip=1truecm
%
The quantum mechanics, and more generally the wave mechanics of
systems that are classically chaotic have drawn much interest
lately; this field is loosely indicated as ``quantum chaos''
 or ``wave chaos''
\cite{HaakeBook,StockmannBook,Dingjan02a,Pechukas84a,Alt97a,Wilkinson01a,Gmachl98}.
Practical experimental systems that display wave chaos are rare;
best known is the 2D microwave stadium-like resonator which has
developed into a very useful tool to study issues of wave chaos
\cite{StockmannBook,Alt97a}. Our interest is in an {\em optical}
implementation of all this; that would allow to study the quantum
aspects of a chaotic wave field, such as random lasing, excess
noise, localization and entanglement
\cite{Beenakker00,Hackenbroich01,Misirpashaev98}.

However, the construction of a high-quality closed resonator (such
as a stadium) is presently impossible in the optical domain due to
the lack of omnidirectional mirror coatings with $R = 100\%$
reflectivity. The best one can do is to use a metal coating;
however this has only $R < 95\%$ in the visible spectrum
\cite{WhiteBook,Wilkinson01a}. Of course, dielectric multilayered
mirrors can reach $R = 99.999\%$ (or more) but these are far from
being omnidirectional.

This leads to the consideration of an {\em open} optical cavity.
One approach is to use a dielectric or semiconductor
microresonator with a deformed cross section and profit from
(non-omnidirectional) total internal reflection
\cite{Gmachl98,Fukushima97}; however, such a microresonator is
difficult to fabricate and control. Our approach is to construct
an open cavity based upon standard high-reflectivity laser
mirrors. We will show, surprisingly, that this open cavity allows
to generate hard chaos; a closed cavity is not required for that.
We will limit ourselves to prove that our system is classically
chaotic; by definition, this is sufficient for a system to be wave
chaotic. A proper wave-mechanical treatment, including the
calculation of the spectrum, will be given later \cite{Aiello03}.

\begin{figure}[!h]
\includegraphics[angle=0,width=7.5truecm]{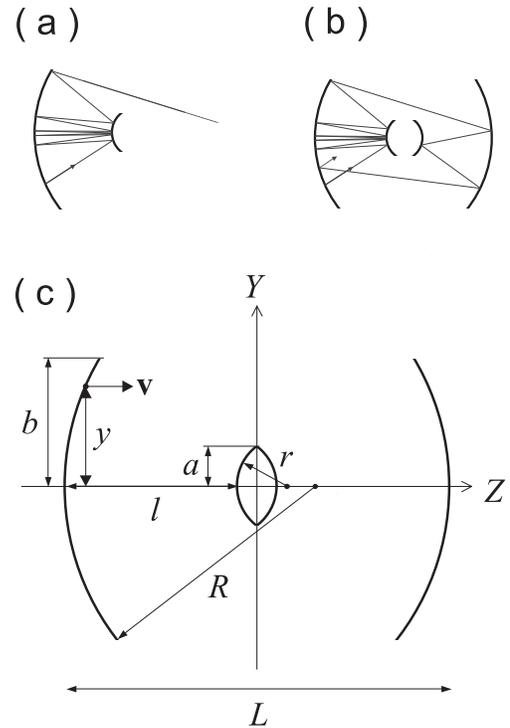}
\caption{\label{fig:1} (a) An unstable cavity is built of a
spherical concave mirror (with radius $R$) and a cylindrical
convex one (with radius $r$) at distance $l < R -r$. (b) Two
unstable cavities are coupled to form a single cavity which is
globally stable for $L < 2 R$. (c) Schematic diagram of the cavity
model utilized in the numerical analysis. For {\em all}
calculations in this Letter we have assumed $a = 0.003$ and $b =
0.025$ where all length are expressed in meters \cite{note1}. On
the left mirror the position $y$ and the velocity $\mathbf{v}$
define the geometry of the chaotic scattering.}
\end{figure}

 The basic idea is as follows (Fig. 1a). An {\em unstable} optical cavity
 \cite{SiegmanBook}
can be built with a concave (focussing) mirror with radius of
curvature $R$ and a convex (dispersing) mirror with radius of
curvature $r$ at distance $l < R -r$. This unstable cavity has
exponential sensitivity to initial conditions \cite{OttBook} but
does not have mixing properties  because an escaping ray never
comes back; therefore chaos cannot occur.
 We overcome this difficulty in the way
illustrated in Fig. 1b: a second cavity, mirror-symmetric to the
first, is utilized to recollect rays leaving the first cavity and
eventually put them back near the starting point. The final design
of our composite cavity is depicted in Fig. 1c; $L$ is the total
length of the cavity and satisfies the relation $L<2R$ in order to
assure the geometrical stability of the whole system. Depending on
the values of $R$, $r$ and $l$ each of the two sub-cavities  can
be either stable or unstable. For $l < R - r$ they are unstable:
this case is the first  object  of our study (see below for  the
case $R- r< l < R$ where the two sub-cavities are stable).
A generic ray lying in the plane of Fig. 1c and undergoing
specular reflections on the cavity mirrors, will never leave this
plane. As will be shown in this Letter, the 2D ray dynamics in
this plane can be completely chaotic; consequently, from now on,
we restrict our attention to the truly 2D cavity shown in Fig. 1c.

The study of the chaotic properties of our composite cavity,
starts from the analogy between {\em geometric optics} of a light
ray and {\em Hamiltonian mechanics} of a point particle
\cite{ArnoldBook}. In this spirit we consider the light ray as a
unit-mass point particle that  undergoes elastic collisions on
hard walls coincident with the surfaces of the mirrors. Between
two consecutive collisions, the motion of the point particle is
determined by the free Hamiltonian $H = \mathbf{p}^2/2$ whereas at
a collision the position $\mathbf{r}(t)$ and the velocity
$\mathbf{v}(t)$ ($|\mathbf{v}(t)| = 1 \; m/s$ throughout this
Letter) of the particle satisfy the law of reflection:
\begin{equation}\label{eq10}
\mathbf{r}(t_+) = \mathbf{r}(t_-), \qquad \mathbf{v}(t_+) =
(\mathbf{1} - 2\mathbf{n}\mathbf{n} )\mathbf{v}(t_-),
\end{equation}
where $t_{\pm} = t \pm 0^{+}$, $\mathbf{n}$ is the unit vector
orthogonal to the surface of the mirror at the point of impact and
the second rank tensor $\mathbf{n}\mathbf{n}$  has Cartesian
components $[\mathbf{n}\mathbf{n}]_{ij} = n_i n_j$ $(i,j = 1,2)$.
The dynamics described by Eqs.(\ref{eq10}) preserves both the
phase space volumes and the symplectic property \cite{OttBook}.
The losses of a cavity due to finite reflectivity of the mirrors
can be quantified by the finesse of the cavity, that is  the
number of bounces that leads to $e^{-1}$ energy decay. For optical
cavities realized with commercially available mirrors values for
the finesse of $10^5$ (or even larger) can be achieved
\cite{note1}. In our model we assume a mirror reflectivity equal
to $100\%$ for all mirrors and consequently we restrict ourselves
to losses  due to the finite transverse dimensions of the concave
mirrors (parameter $b$ in Fig. 1c). {\em If} chaos occurs, almost
all trajectories will escape in the end \cite{OttBook}; the key
question is whether a typical trajectory will survive sufficiently
long that chaos is still a useful concept.

\begin{figure}[!h]
\includegraphics[angle=0,width=8truecm]{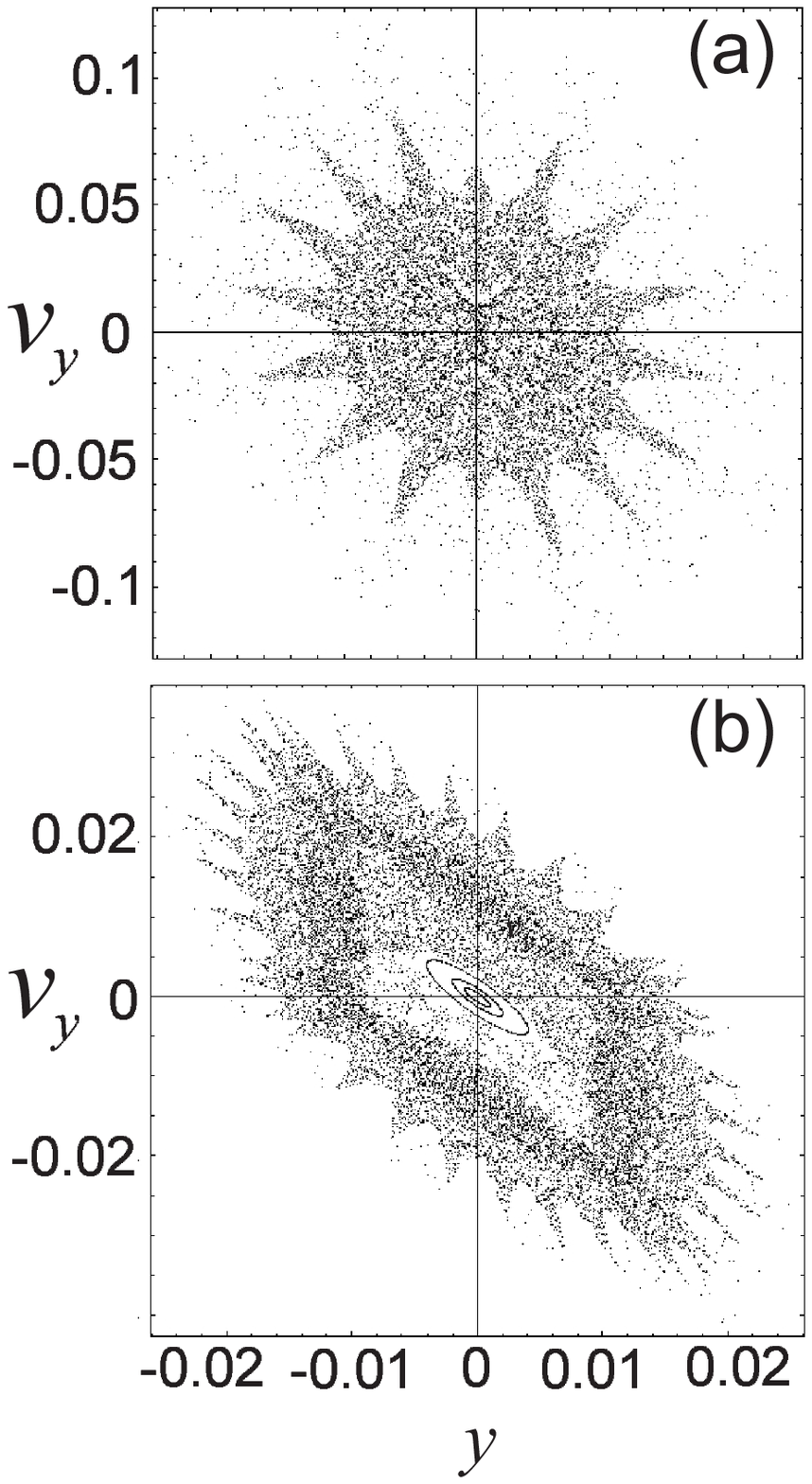}
\caption{\label{fig:2}  Poincar\'{e} maps obtained as described in
the text. In both cases the cavity parameters $a$ and $b$ are
chosen as: $a = 0.003$, $b = 0.025$. Note that $v_y \simeq 0.1$
corresponds to an angle of incidence of about $6^\circ$
\cite{note1}. (a) Hard-chaos configuration: $R = 1$, $r = 0.25$,
$l = 0.04$. A {\em single} chaotic orbit is shown. (b) Soft-chaos
configuration: $R = 1$, $r = 0.9$, $l = 0.45$. Three stable orbits
(concentric ellipses) and a single chaotic one are shown.}
\end{figure}

We use the Poincar\'{e} surface of section (SOS) \cite{OttBook} as
a tool to display the dynamical properties of our composite
cavity. There are several possibilities for defining a SOS; we
choose as reference  surface the left mirror, plotting $y$ and
$v_y$ each time the ray is reflected by that mirror.
 In Fig. 2a we show the SOS generated by a {\em single}
orbit for a cavity configuration such that $l < R - r$
(geometrically unstable sub-cavities). Apparently, hard chaos
occurs; the unstable periodic orbit bouncing along the $Z-$axis of
the cavity is represented by an hyperbolic fixed point on the SOS.
The explicit value of the Lyapunov exponent for this orbit can be
easily calculated exploiting the above mentioned analogy between
geometric optics and Hamiltonian mechanics. We recall that the
magnification $M$ \cite{SiegmanBook} of the cavity shown in Fig.
1a can be easily calculated in terms of $m$, the half of trace of
the $ABCD$ matrix of the cavity:
\begin{equation}\label{eq20}
  M = m + (m^2-1)^{1/2},
  \end{equation}
  where
  \begin{equation}\label{eq25}
m =
  2\left(1-\frac{l}{R}\right)\left(1+\frac{l}{r}\right)-1.
\end{equation}
Since the $ABCD$ matrix of the cavity coincides with the monodromy
matrix \cite{StockmannBook} for the unstable periodic orbit
bouncing back and forth along the $Z$-axis, the positive Lyapunov
exponent $\lambda_0$ for such orbit is given by:
\begin{equation}\label{eq30}
\lambda_0 = \frac{v}{2l}\ln M, \qquad (v = 1 m/s ).
\end{equation}
With the numerical values utilized for Fig. 2a we obtain $M \simeq
1.94$ and $\lambda_0 \simeq 8.27$ (in units of $s^{-1}$).

We may now ask what will happen if the two sub-cavities of Fig. 1c
are {\em stable} (i.e. $R-r<l<R$) so that there is {\em no}
magnification. Will the chaoticity (partly) survive, or not? As
shown in Fig. 2b we find that in that case the periodic orbit
bouncing along the $Z-$axis is  represented by an {\em elliptical}
fixed point on the SOS,
 surrounded by a KAM island of stability in which three stable trajectories are
clearly visible. Surprisingly, we find that despite the stability
of the two sub-cavities, the KAM island of stability is embedded
in a sea of chaotic trajectories with a positive Lyapunov
exponent. Therefore depending on the values of the parameters $R$,
$r$ and $l$ our cavity can exhibit either fully chaotic behavior,
or soft-chaotic behavior with coexistence of ordered and
stochastic trajectories.
\begin{table}
\caption{\label{ta_1} Results for Lyapunov exponents $\lambda_1$
for different cavity configurations which are specified by $l_R$
($l_L$) as the length of the right (left) sub-cavity and $R$ ($r$)
as the radius of curvature of the concave (convex) mirror. All
lengths are given in m, further details are given in the text.}
\begin{tabular}{|c|c|c|c|c|c|}
  % after \\: \hline or \cline{col1-col2} \cline{col3-col4} ...
  \hline
    Configuration & $R $  & $r$ & $l_L$ & $l_R$ & $\lambda_1$ \\
     \hline \hline
  $UU$ & 1 & .25 & .04 & .04 & $ 0.104 \pm 0.004 $ \\
  $US$ & 1 & .90 & .05 & .30 & $ 0.0165 \pm 0.0002$ \\
  $SS$ & 1 & .90 & .45 & .45 & $ 0.0040 \pm 0.0001$ \\
  $MM$ & 1 & .80 & .20 & .20 & $ 0.0090 \pm 0.00015$ \\
  $SM$ & 1 & .80 & .40 & .20 & $ 0.0096 \pm 0.0001 $\\
  $UM$ & 1 & .80 & .01 & .20 & $ 0.0191 \pm 0.0004$ \\ \hline
\end{tabular}
\end{table}
%
%%%%%%%%%%%%%
 A rigorous theory for the
calculation of  average Lyapunov exponents, entropies and escape
time for open systems, has been developed by Gaspard and coworkers
in the last decade \cite{GaspardBook}. In short, in chaotic open
systems there exists a fractal set of never-escaping orbits, the
so-called {\em repeller} \cite{Kadanoff84}
% which constitutes the support of the invariant probability measure
on which quantities as the average Lyapunov exponents can be
evaluated. In a Hamiltonian system with two degrees of freedom the
Lyapunov exponents come in pairs $(\lambda_i, \lambda_{-i})$,
$(i=1,2)$ with $\lambda_1 \geq \lambda_2 = 0$ and $\lambda_i+
\lambda_{-i}=0$ \cite{OttBook}.
We have calculated \cite{Benettin78a,Dellago95b} the values of
$\lambda_1$ for different long-living trajectories belonging to
the repeller for different cavity configurations; the ``pairs
rule" (sum of {\em all} Lyapunov exponents equal to 0) has been
confirmed within our numerical accuracy. The results are shown in
Table I. Each cavity configuration is labelled as $LR$ where $L,R
= U, M, S$ are labels which indicate the stability properties of
the left and right sub-cavity respectively: $U = $ Unstable
($m>1$), $M = $ Marginally stable ($m = 1$), $S =$ Stable ($m
<1$). In all cases we find a positive Lyapunov exponent which
confirms that chaos has developed; the maximum value ($\lambda_1 =
0.104$) occurs for the $UU$ case. We stress the fact that the
values for the parameters characterizing the different cavity
configurations given in Table I are experimentally realistic (i.e.
such mirrors are commercially available).

Note that the   value $\lambda_0 \simeq 8.27$ quoted above for the
{\em periodic} orbit bouncing along the $Z-$axis of the unstable
cavity is quite different from the average value $\lambda_{UU}
\simeq 0.10$. We argue that this is due to the fact that the {\em
same} unstable periodic orbit may be considered as an orbit of the
single half-cavity (very open system: no chaos at all) or as an
orbit of the overall composite cavity. Consequently the sub-cavity
value for $\lambda_1$, even though it remains obviously the same,
is ``diluted'' into the value for the whole cavity.

\begin{figure}[!h]
\includegraphics[angle=0,width=8.5truecm]{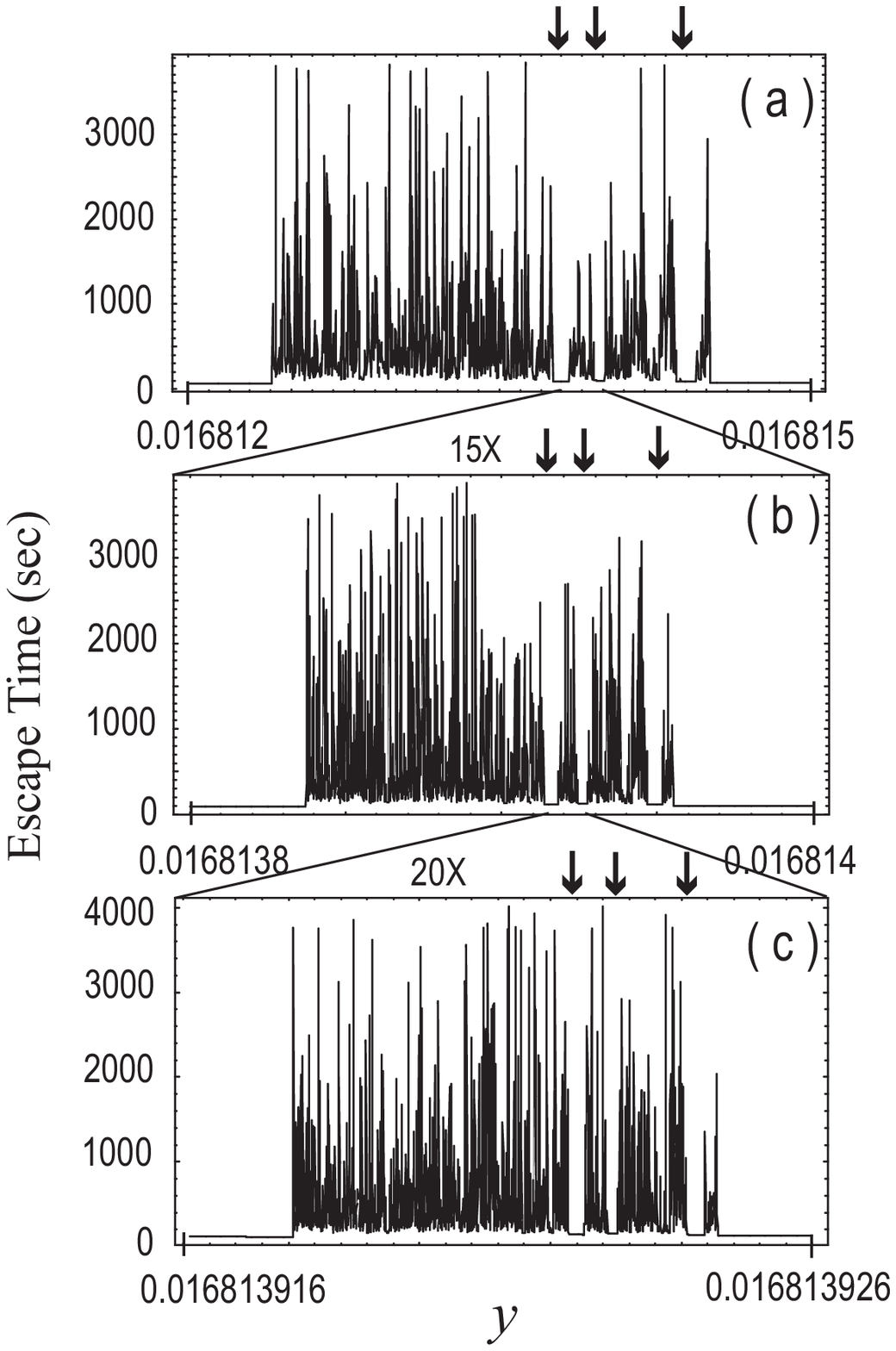}
\caption{\label{fig:3} Escape-time as a function of the impact
parameter $y$ for the same hard-chaos cavity configuration
utilized in Fig. 2a, shown at different relative magnifications
($1 \times$, $15 \times$ and $20 \times$) of the horizontal axis.
The finite value ($\approx 4000$ secs) for each singular peak
corresponds to the finite value ($6 \times 10^4$) of the maximum
number of bounces allowed in the simulations. In the three
pictures are clearly visible the ``windows of continuity''
(arrows) and the repetition of the same pattern on different
scales (self similarity).}
\end{figure}

It is possible to obtain a picture of the repeller set  by
plotting the escape-time function \cite{OttBook}.
 For each trajectory starting at
position $y$ (``impact parameter'') and horizontal velocity
$\mathbf{v}$ on the left mirror (see Fig. 1c), we calculate the
time at which the last bounce occurs before the ray leaves the
cavity. Trajectories which escape in a finite time (almost all),
give a finite value for the escape-time function, whereas trapped
orbits are represented as singular points. By definition, all
initial conditions leading to a singularity of the escape-time
function belong to the repeller itself.
 In Fig. 3 we
show a portion of the  escape time function for a hard-chaos
cavity configuration ($UU$: same parameters as in Fig. 2a). In
Fig. 3a we observe clearly three ``windows of continuity''
\cite{GaspardBook} (arrows) for which the escape time function has
a small value. Actually,  Fig. 3a as a whole results from the
blow-up of the escape time function in a region bounded by two
other windows of continuity which are partially visible on the
left and the right side of the figure. Consecutive blow-ups of
Fig. 3a are shown in Fig. 3b  and Fig. 3c. We note that going from
one picture to the next  both the density and the height of the
singular peaks  increase thus indicating that the repeller set is
dense. Moreover the windows of continuity act as convenient
markers of the self-similar nature of the pattern as a whole; this
self-similarity strongly suggests that the escape  time function
is singular on a fractal set, as expected for a repeller. The
dense occurrence of singular points is a clear signature of the
mixing mechanism due to the  confinement generated by the outer
concave mirrors. Note that typical escape times are much larger
than $1/\lambda_{UU}$, thus allowing ample time for chaos to
develop.

Thus, we have shown that it is possible to build, with
commercially available optical elements,  a composite optical
cavity which displays classical chaotic properties.
%%%%%%%%%%%%%%%%%%%%%%%%%%%%%%
Despite  the ``local'' Hamiltonian structure of its phase space
our optical cavity is, as we had anticipated, an {\em open}
system. Opening up a closed chaotic Hamiltonian system may
generate transiently chaotic behavior due to the escape of almost
all the trajectories \cite{Schneider02a} and our composite cavity
promises an easy experimental realization thereof. Evidence for
ray chaos comes both from the computation of Lyapunov exponents
and from the plot of the escape time functions. The huge density
of singular points in the escape time functions is a strong
indication that the anticipated mixing works: unstable orbits
which leave one half-cavity are recollected by the other
half-cavity until they become again unstable and come back to the
first cavity. Orbits for which this process is repeated forever
generate singularities in the escape function.

Finally,  the most convenient experimental way to realize the
 composite cavity as a 3D system seems to be by using spherical elements
for focussing mirrors and  a cylindrical element for the
dispersing one. A bi-convex cylindrical mirror (with $X$ as the
axis of the cylinder) is dispersing for trajectories lying in a
plane orthogonal to the $X$-axis  and is neutral (flat surface)
for trajectories in a plane containing the $X$-axis itself. In
this case Fig. 1a represents the unstable cross section of the
left sub-cavity.
%%%%%%%%%
The realization of such an open chaotic cavity in the optical
domain opens a broad perspective: many quantum-optics experiments
can now be done on a practical chaotic system
\cite{Beenakker00,Misirpashaev98}. Such experiments will greatly
benefit from the ease of manipulation and control offered by the
macroscopic nature of our composite cavity: work along these lines
is in progress in our group.

This project is part of the program of  FOM and is also supported
by the EU under the IST-ATESIT contract. We acknowledge J. Dingjan
and T. Klaassen for stimulating discussions.

\bibliography{PRL}

\begin{thebibliography}{22}
\bibitem[{not()}]{note1}
\bibinfo{note}{With values for $a$, $b$, $l$, $R$ and $r$ as chosen in Figs.
  1,2, the maximum angle of incidence on the mirrors occurring in Fig. 2 is
  $6^\circ$; this is sufficiently small as not to compromise the high
  reflectivity of a multilayer dielectric mirror}.
\end{thebibliography}

\end{document}